\documentclass{aastex}
\usepackage{spr-astr-addons}
\usepackage{url}

\RequirePackage{color}

\begin{document}

\title{Constraining the degree of the dominant mode in QQ Vir}
\slugcomment{}
%% Running heads
\shorttitle{Mode degree in QQ Vir}
\shortauthors{A.Baran et al.}

\author{Andrzej Baran\altaffilmark{1}}
\affil{Cracow Pedagogical University, Iowa State University}
\and
\author{John Telting\altaffilmark{2}}
\affil{Nordic Optical Telescope, La Palma, Spain}
\and
\author{Roy {\O}stensen\altaffilmark{3}}
\affil{Institute for Astronomy, K.\,U.\,Leuven}
\and
\author{Maciej Winiarski\altaffilmark{1}}
\affil{Cracow Pedagogical University}
\and
\author{Marek Dro\.zd\.z\altaffilmark{1}}
\affil{Cracow Pedagogical University}
\and
\author{Dorota Kozie{\l}\altaffilmark{4}}
\affil{Jagiellonian University}
\and
\author{Mike Reed\altaffilmark{5}}
\affil{Missouri State University}
\and
\author{Raquel Oreiro\altaffilmark{3}}
\affil{Institute for Astronomy, K.\,U.\,Leuven}
\and
\author{Roberto Silvotti\altaffilmark{6}}
\affil{INAF\,--\,Osservatorio Astronomico di Torino}
\and
\author{Micha{\l} Siwak\altaffilmark{4}}
\affil{Jagiellonian University}
\and
\author{Uli Heber\altaffilmark{7}}
\affil{Dr.\,Remeis\,--\,Sternwarte, Institute for Astronomy, University Erlangen\,--\,N\"urnberg}
\and
\author{Peter Papics\altaffilmark{8}}
\affil{Konkoly Observatory}
%\email{\emaila}

%\altaffiltext{1}{First Alternate Affiliation.}
%\altaffiltext{2}{Second Alternate Affilation.}
%\altaffiltext{3}{Third Alternate Affilation.}

\begin{abstract}
We present early results of the application of a method which uses multicolor photometry and spectroscopy for $\ell$ discrimination. This method has been successfully applied to the pulsating hot subdwarf Balloon\,090100001. Here we apply the method to QQ\,Vir (PG1325+101). This star was observed spectroscopically and photometrically in 2008. Details on spectroscopy can be found in Telting et al. (this volume) while photometry and preliminary results on $\ell$ discrimination are provided here. The main aim of this work was to compare the value of the $\ell$ parameter derived for the main mode in QQ\,Vir to previously published values derived by using different methods.
\end{abstract}


%\section*{}
%\label{sec:intro}

\section{The method}
The method we used for mode degree $\ell$ discrimination has been already described and the results of its application to the main sequence stars presented in several papers. This method is very challenging, since it requires observations performed in at least three filters. Moreover, time-series spectroscopy could also be employed. For this reason, the best objects to apply this method to are bright stars with relatively long periods and relatively high amplitudes, so they are accessible even with small size telescopes. It is possible then to achieve good sampling of the brightness changes to derive amplitudes and periods of the detected modes.
Although hot subdwarfs (sdB) are relatively faint objects, in 2004 independent photometric and spectroscopic observations of the brightest pulsating sdB star, Balloon\,090910001 (Bal09), allowed a successful application of the method \citep{baran08}. For the same star, but with high precision multicolor photometry, \cite{charp08} also derived good discrimination for a few modes with highest amplitudes. From these two papers we can draw one important conclusion. The method may give positive discrimination of the $\ell$ parameter either having high precision three\,--\,filter photometry, or a few (at least two) of low precision photometry supported by spectroscopy.
Encouraged by the successful application of the method for Bal09 we decided to try with another pulsating subdwarf B star, QQ\,Vir.

\begin{figure}[]
%\plotone{fig1.ps}
\includegraphics[angle=-90,scale=0.3,width=\columnwidth]{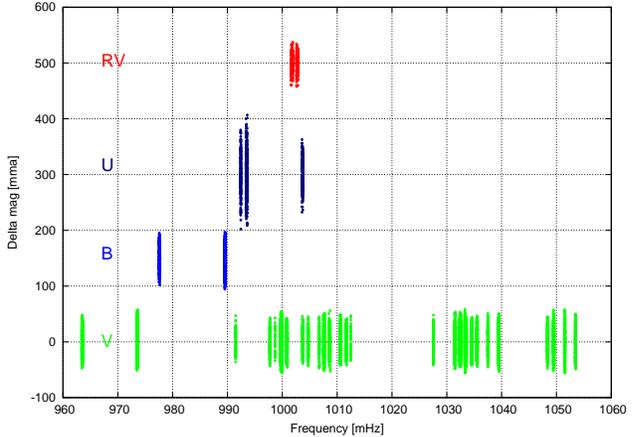}
\caption{Light curve of QQ\,Vir in three filters. Data in different filters are colored according to the bandpass and denoted on the left side. Red data are for the radial velocity curve and the scale on Y-axis is not relevant for these data} %% no full stop at the end
\label{fig1}
\end{figure}

The main goal of performing another observation of QQ\,Vir was to decide which (or if any) value of the $\ell$ parameter for the dominant mode, obtained in previous works, is the correct one. \cite{telt04} analyzed spectroscopic data and stated that the main mode is consistent with $\ell$\,=\,0. Quite interestingly, \cite{charp06}, based on photometry data only, derived $\ell$\,=\,2 from frequency matching. The \cite{charp06} result is even more surprising as it indicates unusually high intrinsic amplitude of the mode with $\ell$\,=\,2 mode.  It is interesting then to use both kinds of data, photometry and spectroscopy, to constrain the value of the mode degree for the dominant mode in QQ\,Vir and if possible, either to confirm one of the already obtained values or to exclude both of them.

\begin{figure}[]
%\plotone{fig2.ps}
\includegraphics[angle=-90,scale=0.3,width=\columnwidth]{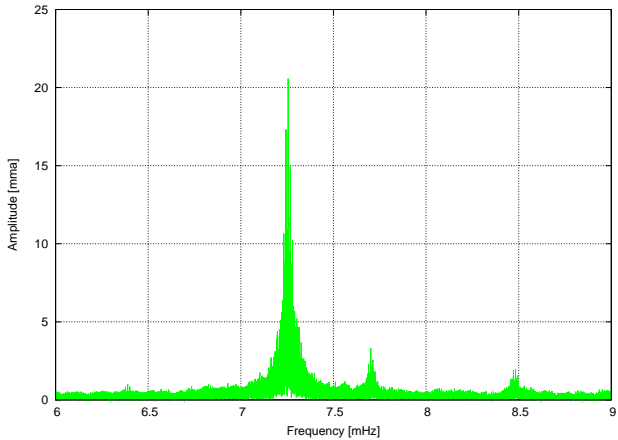}
\caption{Amplitude spectrum of QQ\,Vir in V filter. The dominant mode is clearly seen at $\sim$7.25\,mHz. Significant signal is also present at higher frequencies. This spectrum is constrained to the region where the dominant signal appears} %% no full stop at the end
\label{fig2}
\end{figure}

\section {Observations}
From the previous observations we knew that the pulsational period of the dominant mode is relatively short ($\sim$2\,min.). Along with a low brightness ($\sim$13.5\,mag in the V filter), data required for the method were not easy to obtain, particularly with small size telescopes. For this reason, we performed only one filter photometry on each telescope with the exposure time short enough to get appropriate sampling but not necessarily long enough to have high precision data. Details on the observations are presented in Table\,\ref{tab1}, while all photometry gathered in this project is plotted in Fig.\ref{fig1}. This figure also includes spectroscopic data which have been used in this method. Details on spectroscopy can be found in the paper by \cite{telt10} in this volume.

Data in different filters were taken at different telescopes. This is the reason why there are not the same number of nights for all filters. With respect to the spectrum of the star and spectral response of the CCD we used, the highest signal has been obtained in the V filter. These data have been used for precise frequency determination of the dominant mode. That was also the reason why observations in the V filter have been performed for the longest time from the sites with easy access, like: Suhora, Mercator and Baker. In addition to the V filter, data in the U filter were taken at Loiano and Nordic Optical Telescope (NOT) during three nights in total. Fortunately, the dominant mode is clearly resolved with only one night of data. Sometime later, it turned out that, quite luckily, almost at the same time as our campaign, absolutely independent data in B filter were taken during the course of a different project. That enables us to use three filter data. The B filter data were not crucial, but this method of mode discrimination improves with the number of filters used.

Although the observations in different filters were not conducted simultaneously, we can use published results to see if the amplitude of the dominant frequency changes with time. Results obtained by \cite{telt04} and \cite{silv06} have radial velocity and B filter amplitudes consistent with ours. Those observations were performed independently and at differing times during 2003. We therefore conclude that the amplitudes are not changing significantly and remain stable to within the errors. Moreover, results from the multisite campaign of \cite{silv06} indicate there are no nearby frequencies to the dominant one, even at very low amplitudes. This suggests that our data do not suffer from amplitude variability or the effects of nearby frequencies.

\begin{table}[t]
\caption{Observational log for QQ\,Vir data. All data were taken in 2008.}
\begin{tabular}{ccccc}
\hline
Date   & Site        & Exp [s]     & Length [h] & Filter \\
\hline
\hline
16 Mar & Loiano      & 20          & 2.1       & U \\
17 Mar & Loiano      & 20          & 5.8       & U \\
28 Mar & NOT         & 22          & 2.5       & U \\
\hline
01 Mar & Loiano      & 15          & 4.4       & B \\
13 Mar & Konkoly     & 15          & 4.8       & B \\
\hline
16 Feb & Suhora      & 15          & 4.2       & V \\
26 Feb & Suhora      & 15          & 0.4       & V \\
15 Mar & Mercator    & 20          & 0.8       & V \\
21 Mar & Mercator    & 20          & 1.9       & V \\
22 Mar & Mercator    & 20          & 1.1       & V \\
24 Mar & Baker       & 25          & 8.1       & V \\
25 Mar & Baker       & 30          & 7.5       & V \\
27 Mar & Mercator    & 20          & 2.6       & V \\
28 Mar & Mercator    & 20          & 2.6       & V \\
30 Mar & Mercator    & 20          & 2.2       & V \\
31 Mar & Mercator    & 20          & 7.0       & V \\
01 Apr & Mercator    & 20          & 6.8       & V \\
03 Apr & Mercator    & 20          & 5.9       & V \\
04 Apr & Mercator    & 20          & 4.4       & V \\
05 Apr & Mercator    & 20          & 0.8       & V \\
20 Apr & Suhora      & 15          & 1.3       & V \\
24 Apr & Suhora      & 15          & 4.9       & V \\
25 Apr & Suhora      & 15          & 7.3       & V \\
26 Apr & Suhora      & 15          & 7.3       & V \\
27 Apr & Suhora      & 15          & 4.3       & V \\
28 Apr & Suhora      & 15          & 3.9       & V \\
30 Apr & Suhora      & 15          & 1.2       & V \\
02 May & Suhora      & 15          & 5.1       & V \\
11 May & Suhora      & 15          & 3.8       & V \\
12 May & Suhora      & 15          & 4.9       & V \\
14 May & Suhora      & 15          & 2.8       & V \\
16 May & Suhora      & 15          & 2.3       & V \\
\hline
\end{tabular}
\label{tab1}
\end{table}

\begin{table*}[]
\caption{Results of the prewhitening process. Only results for the main mode at 7.2554776\,mHz are presented. Phases are given for epoch 2454552.0}
\begin{tabular}{ccccc}
\hline
filter & Amplitude [mma] & error [mma] & phase [rad] & error [rad] \\
\hline
U  & 31.89 & 1.02 & 3.425 & 0.048 \\
B  & 26.15 & 0.42 & 3.438 & 0.049 \\
V  & 23.73 & 0.14 & 3.434 & 0.007 \\
\hline
\hline
   & Radial Velocity [km/s] & error [km/s] & phase [rad] & error [rad] \\
\hline
RV & 11.74 & 0.55 & 5.633 & 0.052 \\
\hline
\end{tabular}
\label{tab2}
\end{table*}

\section{Fourier analysis}
A prewhitening procedure was applied to derive frequencies, amplitudes and phases of the pulsation modes. First, we assumed that amplitudes can vary linearly. As it turned out, the change was insignificant (within the errors) and we have decided to assume constant amplitude and redid the prewhitening process. Phases were calculated relative to epoch 2454552.0 which was chosen arbitrarily. Amplitudes and phases for the dominant mode at frequency 7.2554776 $\pm$ 0.0000004\,mHz (derived from V filter data) derived from the data in U B V filters, as well as from the radial velocity curve are presented in Table\,\ref{tab2}. A raw amplitude spectrum calculated from the data in V filter is shown in Fig.\ref{fig2}. It has been limited to the region where the signal appeared, which is 6\,--\,9\,mHz. The amplitude spectra calculated from other filters looks similar, however the noise level and resolution is worse as less data were obtained. Since the main aim of this work is to derive the most likely value of the dominant mode, the other modes are not included in Table\,\ref{tab2}. Nevertheless, during the prewhitening process they were taken into account to check if their removal has any influence on the frequency and/or amplitude of the mode in question. If so, they were included in the solution.

\section{Application of the method}
Having derived frequencies, amplitudes and phases we could apply the method to discriminate $\ell$ parameters for the dominant modes detected both in the radial-velocity and brightness variations.
Discrimination of $\ell$, does not depend on the observations alone. It also depends on the model parameters, in particular effective temperature, surface gravity, flux and limb darkening derivatives over effective temperature and surface gravity. For QQ\,Vir we assumed the effective temperature and surface gravity derived by \cite{telt04}, T$_{\rm eff}$ = 34\,800~K, $\log g$ = 5.81~dex. Then we needed to calculate flux distributions in the range between 300 and 700~nm for a grid of models with T$_{\rm eff}$ ranging from 32\,500 to 37\,500~K and $\log g$, from 5.5  to 6.0~dex. To do this, we used a grid of models prepared for the interested range in T$_{\rm eff}$ and $\log g$. The resulting flux distributions were multiplied by transmission functions of the filters to get integrated fluxes in the three photometric bands -- $UBV$ -- that we used.

\begin{figure}[]
\plotone{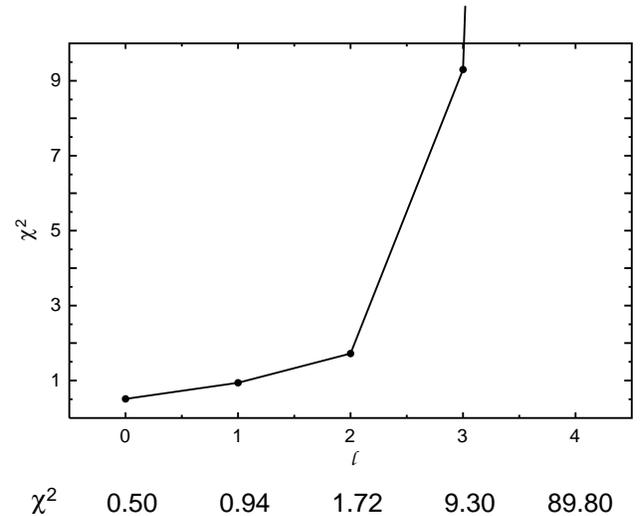}
\caption{Results of application of the method for mode degree discrimination. The plot shows $\chi^2$ as a function of the $\ell$ parameter. In the bottom of the figure the precise values of $\chi^2$ for each $\ell$ are given} %% no full stop at the end
\label{fig3}
\end{figure}

For the same grid of models, we also calculated the specific intensities as a function 
of $\mu$ = $\cos\theta$, where $\theta$ is the angle between the line of sight and normal to the stellar surface in a given point of the stellar disc. Using these intensities, the limb darkening law was derived fitting the coefficients $c$ and $d$ of a function
\begin{equation}
I(\mu)\,=\,I(0)[1-c(1-\mu)-d(1-\sqrt{\mu})]
\label{eq1}
\end{equation}
which was found to reproduce the calculated changes of $I(\mu)$ best. This was done for different combinations of input parameters of the model, thus allowing the calculation of the necessary derivatives.

\section{Discussion and conclusions}
A disadvantage of the method we applied for mode degree discrimination is that we cannot derive a value of $\ell$ in a direct way. We must use an iterative process and employ a merit function to discriminate which mode is the most probable. The minimum of the function (if it exists) indicates the possible value. In this study we applied this method to discriminate between divergent solutions for the main frequency of QQ\,Vir. Results of our calculation are shown in Fig.\ref{fig3}.
The results clearly show that the best fit was obtained for $\ell>$0. However, the minimum of the merit function is not very deep and cannot exclude $\ell$=1 and $\ell$=2. It can, however,, exclude $\ell>$2. This does not help to achieve the main goal of unique mode identification, but is consistent with previous work. In the case of Bal09, \cite{baran08} excluded any $\ell$ value if the $\chi^2$ function was at least 3 times higher than the smallest value of this function. If we follow this rule we can also exclude the $\ell \ge $\,2 value. On the other hand, we can use the probability Q that the merit function for the correct model will exceed a given value by chance. The Q values for the $\ell$=0, 1 and 2 are 0.7, 0.4, 0.13, respectively. In this case none of the first three $\ell$'s can be excluded. However the chance that $\ell$\,=\,0 is wrong is more than 5 times smaller than for $\ell$\,=\,2. 

\acknowledgments
This work was partially supported by Polish MNiSW grant No. 554/MOB/2009/0. RS wishes to thank A.DeBlasi for having contributed to the Loiano observations. R.O. is supported by the Research Council of Leuven University, under grant GOA/2003/04.

%\section{Citing references}

%Use \verb!\cite! command to cite reference(s).
%\smallskip

%\noindent
%\verb!\cite{baran05}! -- \cite{baran05}\\

\nocite{*}
\bibliographystyle{spr-mp-nameyear-cnd}
%\bibliography{myref}
%\bibliography{biblio-u1}

\begin{thebibliography}{}
\bibitem[Baran et al.(2008)]{baran08} Baran, A., Pigulski, A., O'Toole., S., 2008, \mnras, 385, 255
\bibitem[Charpinet et al.(2006)]{charp06} Charpinet, S., Silvotti, R., Bonanno, A., 2006, \aap, 459, 565
\bibitem[Charpinet et al.(2008)]{charp08} Charpinet, S., Fontaine, G., Brassard, P., 2008, Hot subdwarf stars and related objects ASP Conference Series, 392, 297
\bibitem[Silvotti et al.(2006)]{silv06} Silvotti, R., Bonanno, A., Bernabei, S. et al., 2006, \aap, 459, 557
\bibitem[Telting \& {\O}stensen (2004)]{telt04} Telting, J.H. \& {\O}stensen R.H., 2004, \aap, 419, 685
\bibitem[Telting et al.(2010)]{telt10} Telting, J.H. et al., 2010, this volume
\end{thebibliography}

\end{document}